\documentclass[sigconf]{acmart}

\AtBeginDocument{
  }

\copyrightyear{2025}
\acmYear{2025}
\setcopyright{acmlicensed}\acmConference[SIGIR '25]{Proceedings of the 48th International ACM SIGIR Conference on Research and Development in Information Retrieval}{July 13--18, 2025}{Padua, Italy}
\acmBooktitle{Proceedings of the 48th International ACM SIGIR Conference on Research and Development in Information Retrieval (SIGIR '25), July 13--18, 2025, Padua, Italy}
\acmDOI{10.1145/3726302.3730161}
\acmISBN{979-8-4007-1592-1/2025/07}

\begin{document}

\title{AgentCF++: Memory-enhanced LLM-based Agents for Popularity-aware Cross-domain Recommendations}

\author{Jiahao Liu}
\authornote{Equal contribution.}
\orcid{0000-0002-5654-5902}
\author{Shengkang Gu}
\orcid{0009-0006-7033-0162}
\authornotemark[1]
\affiliation{
  \institution{Fudan University}
  \city{Shanghai}
  \country{China}}
\email{jiahaoliu21@m.fudan.edu.cn}
\email{gusk24@m.fudan.edu.cn}

\author{Dongsheng Li}
\orcid{0000-0003-3103-8442}
\affiliation{
  \institution{Microsoft Research Asia}
  \city{Shanghai}
  \country{China}}
\email{dongshengli@fudan.edu.cn}

\author{Guangping Zhang}
\orcid{0009-0001-9853-8268}
\affiliation{
  \institution{Fudan University}
  \city{Shanghai}
  \country{China}}
\email{gpzhang20@fudan.edu.cn}

\author{Mingzhe Han}
\orcid{0000-0002-4911-6093}
\affiliation{
  \institution{Fudan University}
  \city{Shanghai}
  \country{China}}
\email{mzhan22@m.fudan.edu.cn}

\author{Hansu Gu}
\orcid{0000-0002-1426-3210}
\affiliation{
  \institution{Independent}
  \city{Seattle}
  \country{United States}}
\email{hansug@acm.org}

\author{Peng Zhang}
\orcid{0000-0002-9109-4625}
\authornote{Corresponding author.}
\affiliation{
  \institution{Fudan University}
  \city{Shanghai}
  \country{China}}
\email{zhangpeng\_@fudan.edu.cn}

\author{Tun Lu}
\orcid{0000-0002-6633-4826}
\authornotemark[2]
\affiliation{
  \institution{Fudan University}
  \city{Shanghai}
  \country{China}}
\email{lutun@fudan.edu.cn}

\author{Li Shang}
\orcid{0000-0003-3944-7531}
\affiliation{
  \institution{Fudan University}
  \city{Shanghai}
  \country{China}}
\email{lishang@fudan.edu.cn}

\author{Ning Gu}
\orcid{0000-0002-2915-974X}
\affiliation{
  \institution{Fudan University}
  \city{Shanghai}
  \country{China}}
\email{ninggu@fudan.edu.cn}

\renewcommand{\shortauthors}{Jiahao Liu et al.}

\begin{abstract}
LLM-based user agents, which simulate user interaction behavior, are emerging as a promising approach to enhancing recommender systems.
In real-world scenarios, users' interactions often exhibit cross-domain characteristics and are influenced by others.
However, the memory design in current methods causes user agents to introduce significant irrelevant information during decision-making in cross-domain scenarios and makes them unable to recognize the influence of other users' interactions, such as popularity factors.
To tackle this issue, we propose a dual-layer memory architecture combined with a two-step fusion mechanism.
This design avoids irrelevant information during decision-making while ensuring effective integration of cross-domain preferences.
We also introduce the concepts of interest groups and group-shared memory to better capture the influence of popularity factors on users with similar interests.
Comprehensive experiments validate the effectiveness of AgentCF++.
Our code is available at \url{https://github.com/jhliu0807/AgentCF-plus}.
\end{abstract}



\maketitle

\section{Introduction}
\begin{figure*}[t]
  \centering
  \includegraphics[width=0.97\linewidth]{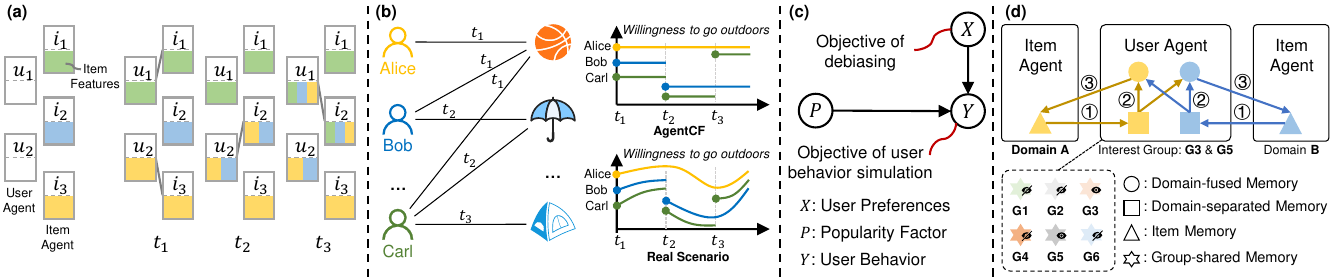}
  \caption{(a) The memory propagation process in AgentCF. (b) An example illustrating AgentCF's limitations in modeling user behavior influenced by popularity factors. (c) Illustration of why modeling popularity factors is necessary for accurately simulating user behavior. (d) Overview of the proposed AgentCF++ model, highlighting its improvements over AgentCF.}
  \label{fig:i9cm}
\end{figure*}

Recommender systems play a pivotal role in the dissemination of information today~\cite{ricci2010introduction,xia2022fire,liu2023personalized,liu2023autoseqrec,liu2023recommendation,liu2023triple,liu2024filtering,liu2022parameter,li2024recommender}.
However, their development is hindered by challenges in effectively understanding user behavior~\cite{shani2011evaluating}.
One promising approach to overcoming these challenges is the reliable simulation of user interaction behavior in a controlled, privacy-preserving manner, thereby improving recommender systems by offering insights into user preferences and system performance~\cite{zhang2024simulating}.
Recent advancements in large language models (LLMs), renowned for their capabilities in understanding, reasoning, and generating content~\cite{zhao2023survey}, have inspired significant efforts to develop LLM-based agents~\cite{wang2024survey}.
These agents often incorporate memory modules~\cite{zhang2024survey}, utilize external tools~\cite{yuan2024easytool}, and engage in advanced reasoning~\cite{huang2024understanding}, enabling them to exhibit emergent human-like behaviors~\cite{park2023generative}.
In this context, researchers have begun investigating the potential of LLM-based agents to simulate user interaction behavior in the field of recommender systems~\cite{wang2024user,zhang2024generative,zhang2024agentcf,shu2024rah,cai2024flow,zhang2024llm}.

Accurately representing user preferences is crucial for a user agent to realistically simulate user behavior.
While various terms are used across studies, this paper uniformly refers to these preferences as being stored in \textit{memory}.
In real-world scenarios, users' interactions with recommender systems often exhibit cross-domain characteristics~\cite{zang2022survey}.
Additionally, individual behaviors are frequently influenced by those of others~\cite{huang2006herding}.
For instance, popularity-related factors suggest that even in the absence of explicit social networks, such influences can be inferred from interaction graphs and propagated through interaction paths~\cite{zhang2021cope}.
However, the current memory design causes the user agent to exhibit two significant limitations:
\textbf{First}, user preferences from multiple domains are mixed into a single memory.
However, only a portion of the preferences is relevant to decision-making in the target domain, leading the user agent to process a considerable amount of irrelevant information.
\textbf{Second}, memory construction relies exclusively on individual user interactions, neglecting how external factors, such as popularity influences, shape user preferences.

In this paper, we present \textbf{AgentCF++}, an enhanced version of AgentCF~\cite{zhang2024agentcf}.
Our approach introduces a \textit{dual-layer memory architecture} comprising \textit{domain-separated memory} and \textit{domain-fused memory}, designed to prioritize target-domain-relevant information in decision-making for cross-domain scenarios.
To refine this architecture, we propose an attention-inspired \textit{two-step fusion mechanism}.
This mechanism first identifies valuable cross-domain knowledge pertinent to the target domain and then integrates these critical preferences.
Furthermore, we introduce the concept of \textit{interest groups} and propose a \textit{group-shared memory mechanism} to facilitate the transfer of popularity effects within the same interest group.
By utilizing interest groups, we ensure that a user’s behavior impacts only those with similar interests, effectively preventing the spread of popularity effects to unrelated users.
Our experimental results on five cross-domain datasets demonstrate the effectiveness of the proposed modules.

\section{Preliminaries}

\subsection{AgentCF}
Unlike previous studies~\cite{wang2024user,zhang2024generative,shu2024rah,cai2024flow,zhang2024llm} that consider only users as agents, AgentCF~\cite{zhang2024agentcf} treats both users and items as agents.
Each user agent is equipped with a memory to store individual preferences, while each item agent maintains a memory to track the interest levels of users with varying preferences towards it.
At each step, leveraging LLMs for decision-making and reflection, these agents autonomously interact, compare their actions with real-world data, and collaboratively adjust their memories to better align with observed behaviors.
As illustrated in Figure~\ref{fig:i9cm}(a), user and item memories gradually propagate over time through interactions, embodying the principles of collaborative filtering.

\subsection{Limitations of AgentCF}
AgentCF employs a single memory for each user agent and item agent.
In cross-domain scenarios, the propagation process integrates information from multiple domains into the memory of each user and item.
On one hand, the mixing of cross-domain preferences in the user agent may introduce noise, thereby complicating decision-making on target domain.
On the other hand, such intermingling may cause the item agent to lose its original domain characteristics.

Additionally, while AgentCF employs collaborative filtering to capture the influence of others' interactions, memory updates for user and item agents occur only during direct interactions, limiting its capacity to comprehensively model how popularity factors shape user behavior.
To illustrate this limitation, consider the following example.
Figure \ref{fig:i9cm}(b) visualizes user-item interaction dynamics using a timestamped bipartite graph.
At $t_1$, Alice, Bob, and Carl purchase outdoor activity items; at $t_2$, Bob and Carl buy rain gear; at $t_3$, Carl purchases camping equipment.
Under the AgentCF framework, Alice, who ceases interactions after $t_1$, is still assumed to engage in outdoor activities, despite worsening weather conditions at $t_2$.
Similarly, Bob shows no interest at $t_3$, failing to account for the improved weather conditions.
This indicates that, within the AgentCF framework, users' memories remain static in the absence of additional interactions.
In practice, however, even without further direct participation, Alice might infer the weather changes at $t_2$ and $t_3$ by observing Bob's and Carl's actions.
This underscores the need for user agents to update their memories not only through their own interactions with item agents but also in response to interactions by other user agents.
In essence, a user’s memory should evolve dynamically, even without direct participation—a critical capability currently missing in the AgentCF framework.

\textbf{Clarification.}
As shown in Figure \ref{fig:i9cm}(c), user behavior ($Y$) is influenced by both user preferences ($X$) and popularity factors ($P$).  
Unlike debiasing approaches, which seek to model user preferences by removing the effects of popularity factors~\cite{ge2024survey,chen2023bias}, user behavior simulation aims to model user behavior.  
Therefore, in user behavior simulation, popularity factors are not a nuisance to be mitigated but a critical factor to be explicitly modeled.

\section{AgentCF++}
Similar to AgentCF, AgentCF++ simulates interactions between user agents and item agents from multiple domains and performs reflection during this process to update the memories of both sides, thereby aligning with user behavior.

\subsection{Memory Architecture}
AgentCF++ employs a similar memory architecture to AgentCF for the item agent, using a single memory to record the interest levels of users with various preferences towards it.
Initially, the item's memory is seeded with its side information.
However, AgentCF++ has meticulously designed the memory architecture for the user agent to enhance its functionality.

Firstly, each user agent in AgentCF++ is equipped with a \textbf{dual-layer memory architecture}.
Specifically, each user agent maintains two types of memory for \textbf{each domain}:
(1) \textbf{Domain-separated memory} retains the user’s preferences specific to a single domain.  
(2) \textbf{Domain-fused memory} also stores preferences within a particular domain but integrates domain-separated memories from other domains.
Initially, both memories are empty.

Additionally, each user agent is assigned to several \textbf{interest groups} through the following process:
(1) \textit{Building user-tag relationships}: The user agent’s domain-fused memory is processed by an LLM to derive a set of interest tags representing the user’s preferences.
(2) \textit{Merging synonymous tags}: The LLM transforms all tags into embedding vectors, which are then grouped into clusters based on semantic similarity using a K-means clustering algorithm.
Each cluster corresponds to a specific area of interest, encompassing tags with high semantic similarity.
(3) \textit{Refining interest groups}: The LLM synthesizes the tags in each cluster to generate a consolidated interest group name.
Ultimately, only the largest few interest groups are retained, collectively covering the majority of the user's interests.
AgentCF++ periodically re-segments the interest groups to ensure they reflect any updates in user preferences.

Each interest group is equipped with a \textbf{group-shared memory}, enabling all user agents within the group to collaboratively access shared information.
The shared memory is of fixed size, designed to store the recent interaction history of its associated users.


\subsection{Inference Phase}
We assume that the current interaction is $(u, i, d)$, where $u$ represents a user agent, $i$ represents an item agent, and $d$ denotes the domain of $i$.
First, a negative sample $j$ is selected from the domain $d$.
Then, $u$ receives the memories of $i$ and $j$ and is tasked with identifying the positive sample while explaining its reasoning.
To mitigate potential position bias, in which LLMs tend to favor earlier options, $j$ is deliberately placed before $i$.
Note that $u$'s decisions depend simultaneously on both domain-separated memory and domain-fused memory within domain $d$, as well as on the group-shared memories it can access.

\textbf{Discussion.}
The dual-layer memory architecture includes a domain-separated memory and a domain-fused memory corresponding to each domain.  
With this memory enhancement, only information relevant to the target domain is utilized during decision-making, effectively reducing noise in cross-domain scenarios.
On the other hand, the sharing mechanism allows user behavior to influence related users without directly updating their individual memories, incorporating the influence of popularity factors into preferences modeling.
For instance, in the scenario depicted in Figure~\ref{fig:i9cm}(b), Bob and Carl insert the behavior of purchasing rain gear into the memory shared with Alice at $t_2$.
At $t_3$, Carl adds the behavior of purchasing camping gear into the memory shared with both Alice and Carl.
This results in Alice’s willingness to go outdoors decreasing at $t_2$ and increasing at time $t_3$, reflecting real-world patterns where user behavior is influenced by trends and popularity factors.
Note that we segment users based on their interests rather than the similarity of their interaction history, to more precisely identify populations influenced by popularity factors.

\subsection{Update Phase}
The memories are updated using a reflection mechanism~\cite{shinn2024reflexion,madaan2024self,pan2023automatically}.
Specifically:
(1) According to the results of the inference phase, $u$ updates its domain-separated memory in domain $d$ using the memories of $i$ and $j$.
This step enables $u$ to learn what it like and dislike from the latest interaction.
(2) We propose a \textbf{two-step fusion mechanism} to effectively integrate information from multiple domains.
Firstly, $u$ extracts preferences related to the target domain $d$ from domain-separated memories of other domains.
Then, $u$ updates its domain-fused memory in domain $d$ based on the extracted preferences.
(3) $i$ and $j$ update their item memories using $u$'s domain-fused memory in domain $d$. 
In this step, $i$ learns which user preferences it appeals to, while $j$ learns which user preferences it does not appeal to.

\textbf{Discussion.}
Figure~\ref{fig:i9cm}(d) illustrates the process of update phase.
The two-step fusion mechanism implicitly incorporates the idea of the attention mechanism, ensuring that the domain-fused memory effectively integrates preferences from other domains while retaining only preferences relevant to the corresponding domain.
Specifically, in the first step, only preferences related to the target domain are extracted, akin to the computation of attention scores.  
In the second step, the extracted preferences from different domains are integrated, akin to the weighted aggregation process in the attention mechanism.
Additionally, with the aid of the reflection mechanism, the cyclic updates of item memory, domain-separated memory, and domain-fused memory enable all memories to achieve self-improvement.

\section{Experiments}

\subsection{Settings}

\subsubsection{Datasets}
We experimented with the Amazon review dataset~\cite{hou2024bridging}.
We constructed the cross-domain datasets \textit{Cross-1}---\textit{Cross-5} by combining data from the Books, CDs, Movies, and Games domains, selecting 3 or 4 domains for each dataset.
Then, we retained interaction records with ratings $\geq 4$ and timestamps spanning six months, from October 2021 to March 2022.
We further filtered the data to include only records of users who interacted across multiple domains and had $\ge$ 10 total interactions.
Following AgentCF, we randomly sampled 100 users to minimize API call expenses.
Next, we sorted these interaction records chronologically and split them into training, validation, and test sets with an 8:1:1 ratio.

\subsubsection{Evaluation}
For each ground truth item, we randomly sample 9 items from the same domain that the user has not interacted with to construct the candidate set.
The user agent is then tasked with ranking these items, and the ranking performance is measured using NDCG and MRR.
We report the average results over 5 runs.

\subsubsection{Baselines}
We used two traditional recommendation models, BPR-MF~\cite{rendle2012bpr} and SASRec~\cite{kang2018self}, as well as four training-free methods, Pop, LLMSeqSim~\cite{harte2023leveraging}, LLMRank~\cite{hou2024large}, and AgentCF~\cite{zhang2024agentcf}, as baseline methods for comparison.
Specifically, Pop ranks candidates based on item popularity, LLMSeqSim evaluates candidates by measuring their similarity to the user's interaction history, and LLMRank employs an LLM as a zero-shot ranker to prioritize candidate items.

We compared solely with AgentCF, omitting other LLM-based user agent methods.
Our goal is to demonstrate that the proposed module enhances AgentCF.
Performance differences with other methods may stem from differing agent construction paradigms, making it challenging to attribute improvements directly to the proposed module.
On the other hand, they all construct memory directly through the user's interaction history, which can, to some extent, be considered equivalent to LLMRank.

We also designed three ablation variants for AgentCF++:
(1) \textit{AgentCF + dual}: Extends AgentCF with only the dual-layer memory architecture.
(2) \textit{AgentCF + shared}: Extends AgentCF with only interest groups and group-shared memory.
(3) \textit{AgentCF++ w/o group}: Users are grouped based on their full interaction history rather than their interests.
The other components, including treating users as agents, treating items as agents, the automatic interaction process, and the reflection mechanism, have already been validated as effective by AgentCF.
Therefore, these components were not included in the ablation study.

\subsection{Overall Performance}
As shown in Table~\ref{tab:d9jv}, on cross-domain datasets, AgentCF achieves comparable performance to training-free methods such as LLMRank but fails to surpass traditional recommendation models like SASRec.
This suggests that traditional models inherently capture popularity factors and cross-domain collaborative information through their training mechanisms, giving them a clear advantage over LLM-based user agents in predicting user behavior.
Encouragingly, the proposed AgentCF++ consistently outperforms both its ablation variants and all baselines. Moreover, the ablation variants consistently outperform AgentCF, further validating the effectiveness of the proposed modules.
Notably, \textit{AgentCF++ w/o group} performs not only worse than AgentCF++ but also worse than \textit{AgentCF + dual}, further underscoring the importance of dividing users into interest groups.
This indicates that assigning users to groups based on their full interaction history is too coarse, allowing the popularity factor to influence unrelated users, introducing noise, and ultimately reducing accuracy.

\begin{table} \footnotesize
  \caption{Overall performance. Due to space constraints, we only report MRR results. NDCG results, available in the repository, lead to consistent conclusions.}
  \label{tab:d9jv}
\begin{tabular}{c|ccccc}
\hline
Method                       & Cross-1         & Cross-2         & Cross-3         & Cross-4         & Cross-5         \\ \hline
BPR-MF                       & 0.2949          & 0.2959          & 0.3114          & 0.3012          & 0.3127          \\
SASRec                       & 0.3463          & 0.3154          & 0.3828          & 0.3118          & 0.3687          \\ \hline
Pop                          & 0.2589          & 0.2817          & 0.3094          & 0.2954          & 0.3089          \\
LLMSeqSim                    & 0.2646          & 0.2549          & 0.3101          & 0.2959          & 0.3124          \\
LLMRank                      & 0.3268          & 0.2730          & 0.3106          & 0.2970          & 0.3308          \\
AgentCF                      & 0.3284          & 0.2681          & 0.3114          & 0.3032          & 0.3480          \\
\textbf{AgentCF++}           & \textbf{0.3537} & \textbf{0.3176} & \textbf{0.3989} & \textbf{0.3321} & \textbf{0.3837} \\ \hline
\textit{AgentCF + dual}      & 0.3495          & 0.2962          & 0.3581          & 0.3139          & 0.3581          \\
\textit{AgentCF + shared}    & 0.3488          & 0.2777          & 0.3190          & 0.3147          & 0.3689          \\
\textit{AgentCF++ w/o group} & 0.3415          & 0.2724          & 0.3181          & 0.3126          & 0.3549          \\ \hline
\end{tabular}
\end{table}





\section{Related Work}
LLM-based agents in recommender systems can be broadly divided into two categories.
The first category focuses on \textbf{recommendation agents} that leverage LLMs to generate or improve recommendations~\cite{shi2024large,zhang2023recommendation,huang2023recommender,wang2023recmind,wang2024multi,zhao2024let,zhang2024prospect}.
%
The second category explores \textbf{user agents} that leverage LLMs to simulate user behavior.
While some studies focus on simulating user dialogues in conversational recommendation~\cite{zhu2024llm,zhu2024reliable,kim2024stop,friedman2023leveraging,wang2023rethinking}, our emphasis is on simulating user interaction behavior.
RecAgent~\cite{wang2024user} and Agent4Rec~\cite{zhang2024generative} employ LLM-based agents, incorporating user profiles, memory, and action modules, to simulate interactions with recommender systems.
RAH~\cite{shu2024rah} places LLM-based multi-agents between users and recommender systems, serving both as recommendation agents and as proxies for user interactions.
FLOW~\cite{cai2024flow} facilitates collaboration between recommendation agents and user agents by establishing a feedback loop.
\citet{zhang2024llm} integrate explicit user preferences, LLM-driven sentiment analysis, a human engagement model, and a statistical framework to robustly simulate user interactions.
AgentCF~\cite{zhang2024agentcf} proposes a novel approach, conceptualizing users and items as agents and employing a collaborative learning strategy to optimize them simultaneously.

Several studies have highlighted LLMs' generalization capabilities in \textbf{cross-domain recommendation}~\cite{bao2023tallrec,bai2024aligning,shen2024exploring,vajjala2024cross,petruzzelli2024instructing,tang2023one} and explored \textbf{popularity bias} in LLM-based recommenders~\cite{jiang2024item,lichtenberg2024large,gao2024sprec,ortega2024evaluating,deldjoo2024understanding}.
These works mainly use LLMs as recommenders, not for simulating user behavior.
Importantly, we emphasize the need to explicitly model popularity factors when simulating user behavior, rather than merely reducing their influence.


\section{Conclusions}
We propose AgentCF++, which consists of:
(1) a dual-layer memory architecture and a two-step fusion mechanism that allow the user agent to avoid introducing irrelevant in cross-domain scenarios;
(2) the concept of interest groups and a shared memory mechanism that captures the influence of popularity among users with similar interests.
Comprehensive experiments demonstrate the effectiveness of AgentCF++.
In the future, we aim to adapt these designs to other LLM-based user agent frameworks.

\begin{acks}
This work is supported by National Natural Science Foundation of China (NSFC) under the Grant No. 62172106. Peng Zhang is a faculty of School of Computer Science, Fudan University. Tun Lu is a faculty of School of Computer Science, Shanghai Key Laboratory of Data Science, Fudan Institute on Aging, MOE Laboratory for National Development and Intelligent Governance, and Shanghai Institute of Intelligent Electronics \& Systems, Fudan University.
\end{acks}

\bibliographystyle{ACM-Reference-Format}
\balance
\bibliography{main}


\begin{thebibliography}{59}


\ifx \showCODEN    \undefined \def \showCODEN     #1{\unskip}     \fi
\ifx \showISBNx    \undefined \def \showISBNx     #1{\unskip}     \fi
\ifx \showISBNxiii \undefined \def \showISBNxiii  #1{\unskip}     \fi
\ifx \showISSN     \undefined \def \showISSN      #1{\unskip}     \fi
\ifx \showLCCN     \undefined \def \showLCCN      #1{\unskip}     \fi
\ifx \shownote     \undefined \def \shownote      #1{#1}          \fi
\ifx \showarticletitle \undefined \def \showarticletitle #1{#1}   \fi
\ifx \showURL      \undefined \def \showURL       {\relax}        \fi
\providecommand\bibfield[2]{#2}
\providecommand\bibinfo[2]{#2}
\providecommand\natexlab[1]{#1}
\providecommand\showeprint[2][]{arXiv:#2}

\bibitem[Bai et~al\mbox{.}(2024)]%
        {bai2024aligning}
\bibfield{author}{\bibinfo{person}{Zhuoxi Bai}, \bibinfo{person}{Ning Wu}, \bibinfo{person}{Fengyu Cai}, \bibinfo{person}{Xinyi Zhu}, {and} \bibinfo{person}{Yun Xiong}.} \bibinfo{year}{2024}\natexlab{}.
\newblock \showarticletitle{Aligning Large Language Model with Direct Multi-Preference Optimization for Recommendation}. In \bibinfo{booktitle}{\emph{Proceedings of the 33rd ACM International Conference on Information and Knowledge Management}}. \bibinfo{pages}{76--86}.
\newblock


\bibitem[Bao et~al\mbox{.}(2023)]%
        {bao2023tallrec}
\bibfield{author}{\bibinfo{person}{Keqin Bao}, \bibinfo{person}{Jizhi Zhang}, \bibinfo{person}{Yang Zhang}, \bibinfo{person}{Wenjie Wang}, \bibinfo{person}{Fuli Feng}, {and} \bibinfo{person}{Xiangnan He}.} \bibinfo{year}{2023}\natexlab{}.
\newblock \showarticletitle{Tallrec: An effective and efficient tuning framework to align large language model with recommendation}. In \bibinfo{booktitle}{\emph{Proceedings of the 17th ACM Conference on Recommender Systems}}. \bibinfo{pages}{1007--1014}.
\newblock


\bibitem[Cai et~al\mbox{.}(2024)]%
        {cai2024flow}
\bibfield{author}{\bibinfo{person}{Shihao Cai}, \bibinfo{person}{Jizhi Zhang}, \bibinfo{person}{Keqin Bao}, \bibinfo{person}{Chongming Gao}, {and} \bibinfo{person}{Fuli Feng}.} \bibinfo{year}{2024}\natexlab{}.
\newblock \showarticletitle{FLOW: A Feedback LOop FrameWork for Simultaneously Enhancing Recommendation and User Agents}.
\newblock \bibinfo{journal}{\emph{arXiv preprint arXiv:2410.20027}} (\bibinfo{year}{2024}).
\newblock


\bibitem[Chen et~al\mbox{.}(2023)]%
        {chen2023bias}
\bibfield{author}{\bibinfo{person}{Jiawei Chen}, \bibinfo{person}{Hande Dong}, \bibinfo{person}{Xiang Wang}, \bibinfo{person}{Fuli Feng}, \bibinfo{person}{Meng Wang}, {and} \bibinfo{person}{Xiangnan He}.} \bibinfo{year}{2023}\natexlab{}.
\newblock \showarticletitle{Bias and debias in recommender system: A survey and future directions}.
\newblock \bibinfo{journal}{\emph{ACM Transactions on Information Systems}} \bibinfo{volume}{41}, \bibinfo{number}{3} (\bibinfo{year}{2023}), \bibinfo{pages}{1--39}.
\newblock


\bibitem[Deldjoo(2024)]%
        {deldjoo2024understanding}
\bibfield{author}{\bibinfo{person}{Yashar Deldjoo}.} \bibinfo{year}{2024}\natexlab{}.
\newblock \showarticletitle{Understanding biases in chatgpt-based recommender systems: Provider fairness, temporal stability, and recency}.
\newblock \bibinfo{journal}{\emph{ACM Transactions on Recommender Systems}} (\bibinfo{year}{2024}).
\newblock


\bibitem[Friedman et~al\mbox{.}(2023)]%
        {friedman2023leveraging}
\bibfield{author}{\bibinfo{person}{Luke Friedman}, \bibinfo{person}{Sameer Ahuja}, \bibinfo{person}{David Allen}, \bibinfo{person}{Zhenning Tan}, \bibinfo{person}{Hakim Sidahmed}, \bibinfo{person}{Changbo Long}, \bibinfo{person}{Jun Xie}, \bibinfo{person}{Gabriel Schubiner}, \bibinfo{person}{Ajay Patel}, \bibinfo{person}{Harsh Lara}, {et~al\mbox{.}}} \bibinfo{year}{2023}\natexlab{}.
\newblock \showarticletitle{Leveraging large language models in conversational recommender systems}.
\newblock \bibinfo{journal}{\emph{arXiv preprint arXiv:2305.07961}} (\bibinfo{year}{2023}).
\newblock


\bibitem[Gao et~al\mbox{.}(2024)]%
        {gao2024sprec}
\bibfield{author}{\bibinfo{person}{Chongming Gao}, \bibinfo{person}{Ruijun Chen}, \bibinfo{person}{Shuai Yuan}, \bibinfo{person}{Kexin Huang}, \bibinfo{person}{Yuanqing Yu}, {and} \bibinfo{person}{Xiangnan He}.} \bibinfo{year}{2024}\natexlab{}.
\newblock \showarticletitle{SPRec: Leveraging Self-Play to Debias Preference Alignment for Large Language Model-based Recommendations}.
\newblock \bibinfo{journal}{\emph{arXiv preprint arXiv:2412.09243}} (\bibinfo{year}{2024}).
\newblock


\bibitem[Ge et~al\mbox{.}(2024)]%
        {ge2024survey}
\bibfield{author}{\bibinfo{person}{Yingqiang Ge}, \bibinfo{person}{Shuchang Liu}, \bibinfo{person}{Zuohui Fu}, \bibinfo{person}{Juntao Tan}, \bibinfo{person}{Zelong Li}, \bibinfo{person}{Shuyuan Xu}, \bibinfo{person}{Yunqi Li}, \bibinfo{person}{Yikun Xian}, {and} \bibinfo{person}{Yongfeng Zhang}.} \bibinfo{year}{2024}\natexlab{}.
\newblock \showarticletitle{A survey on trustworthy recommender systems}.
\newblock \bibinfo{journal}{\emph{ACM Transactions on Recommender Systems}} \bibinfo{volume}{3}, \bibinfo{number}{2} (\bibinfo{year}{2024}), \bibinfo{pages}{1--68}.
\newblock


\bibitem[Harte et~al\mbox{.}(2023)]%
        {harte2023leveraging}
\bibfield{author}{\bibinfo{person}{Jesse Harte}, \bibinfo{person}{Wouter Zorgdrager}, \bibinfo{person}{Panos Louridas}, \bibinfo{person}{Asterios Katsifodimos}, \bibinfo{person}{Dietmar Jannach}, {and} \bibinfo{person}{Marios Fragkoulis}.} \bibinfo{year}{2023}\natexlab{}.
\newblock \showarticletitle{Leveraging large language models for sequential recommendation}. In \bibinfo{booktitle}{\emph{Proceedings of the 17th ACM Conference on Recommender Systems}}. \bibinfo{pages}{1096--1102}.
\newblock


\bibitem[Hou et~al\mbox{.}(2024a)]%
        {hou2024bridging}
\bibfield{author}{\bibinfo{person}{Yupeng Hou}, \bibinfo{person}{Jiacheng Li}, \bibinfo{person}{Zhankui He}, \bibinfo{person}{An Yan}, \bibinfo{person}{Xiusi Chen}, {and} \bibinfo{person}{Julian McAuley}.} \bibinfo{year}{2024}\natexlab{a}.
\newblock \showarticletitle{Bridging language and items for retrieval and recommendation}.
\newblock \bibinfo{journal}{\emph{arXiv preprint arXiv:2403.03952}} (\bibinfo{year}{2024}).
\newblock


\bibitem[Hou et~al\mbox{.}(2024b)]%
        {hou2024large}
\bibfield{author}{\bibinfo{person}{Yupeng Hou}, \bibinfo{person}{Junjie Zhang}, \bibinfo{person}{Zihan Lin}, \bibinfo{person}{Hongyu Lu}, \bibinfo{person}{Ruobing Xie}, \bibinfo{person}{Julian McAuley}, {and} \bibinfo{person}{Wayne~Xin Zhao}.} \bibinfo{year}{2024}\natexlab{b}.
\newblock \showarticletitle{Large language models are zero-shot rankers for recommender systems}. In \bibinfo{booktitle}{\emph{European Conference on Information Retrieval}}. Springer, \bibinfo{pages}{364--381}.
\newblock


\bibitem[Huang and Chen(2006)]%
        {huang2006herding}
\bibfield{author}{\bibinfo{person}{Jen-Hung Huang} {and} \bibinfo{person}{Yi-Fen Chen}.} \bibinfo{year}{2006}\natexlab{}.
\newblock \showarticletitle{Herding in online product choice}.
\newblock \bibinfo{journal}{\emph{Psychology \& Marketing}} \bibinfo{volume}{23}, \bibinfo{number}{5} (\bibinfo{year}{2006}), \bibinfo{pages}{413--428}.
\newblock


\bibitem[Huang et~al\mbox{.}(2023)]%
        {huang2023recommender}
\bibfield{author}{\bibinfo{person}{Xu Huang}, \bibinfo{person}{Jianxun Lian}, \bibinfo{person}{Yuxuan Lei}, \bibinfo{person}{Jing Yao}, \bibinfo{person}{Defu Lian}, {and} \bibinfo{person}{Xing Xie}.} \bibinfo{year}{2023}\natexlab{}.
\newblock \showarticletitle{Recommender ai agent: Integrating large language models for interactive recommendations}.
\newblock \bibinfo{journal}{\emph{arXiv preprint arXiv:2308.16505}} (\bibinfo{year}{2023}).
\newblock


\bibitem[Huang et~al\mbox{.}(2024)]%
        {huang2024understanding}
\bibfield{author}{\bibinfo{person}{Xu Huang}, \bibinfo{person}{Weiwen Liu}, \bibinfo{person}{Xiaolong Chen}, \bibinfo{person}{Xingmei Wang}, \bibinfo{person}{Hao Wang}, \bibinfo{person}{Defu Lian}, \bibinfo{person}{Yasheng Wang}, \bibinfo{person}{Ruiming Tang}, {and} \bibinfo{person}{Enhong Chen}.} \bibinfo{year}{2024}\natexlab{}.
\newblock \showarticletitle{Understanding the planning of LLM agents: A survey}.
\newblock \bibinfo{journal}{\emph{arXiv preprint arXiv:2402.02716}} (\bibinfo{year}{2024}).
\newblock


\bibitem[Jiang et~al\mbox{.}(2024)]%
        {jiang2024item}
\bibfield{author}{\bibinfo{person}{Meng Jiang}, \bibinfo{person}{Keqin Bao}, \bibinfo{person}{Jizhi Zhang}, \bibinfo{person}{Wenjie Wang}, \bibinfo{person}{Zhengyi Yang}, \bibinfo{person}{Fuli Feng}, {and} \bibinfo{person}{Xiangnan He}.} \bibinfo{year}{2024}\natexlab{}.
\newblock \showarticletitle{Item-side Fairness of Large Language Model-based Recommendation System}. In \bibinfo{booktitle}{\emph{Proceedings of the ACM on Web Conference 2024}}. \bibinfo{pages}{4717--4726}.
\newblock


\bibitem[Kang and McAuley(2018)]%
        {kang2018self}
\bibfield{author}{\bibinfo{person}{Wang-Cheng Kang} {and} \bibinfo{person}{Julian McAuley}.} \bibinfo{year}{2018}\natexlab{}.
\newblock \showarticletitle{Self-attentive sequential recommendation}. In \bibinfo{booktitle}{\emph{2018 IEEE international conference on data mining (ICDM)}}. IEEE, \bibinfo{pages}{197--206}.
\newblock


\bibitem[Kim et~al\mbox{.}(2024)]%
        {kim2024stop}
\bibfield{author}{\bibinfo{person}{Sunghwan Kim}, \bibinfo{person}{Tongyoung Kim}, \bibinfo{person}{Kwangwook Seo}, \bibinfo{person}{Jinyoung Yeo}, {and} \bibinfo{person}{Dongha Lee}.} \bibinfo{year}{2024}\natexlab{}.
\newblock \showarticletitle{Stop Playing the Guessing Game! Target-free User Simulation for Evaluating Conversational Recommender Systems}.
\newblock \bibinfo{journal}{\emph{arXiv preprint arXiv:2411.16160}} (\bibinfo{year}{2024}).
\newblock


\bibitem[Li et~al\mbox{.}(2024)]%
        {li2024recommender}
\bibfield{author}{\bibinfo{person}{Dongsheng Li}, \bibinfo{person}{Jianxun Lian}, \bibinfo{person}{Le Zhang}, \bibinfo{person}{Kan Ren}, \bibinfo{person}{Tun Lu}, \bibinfo{person}{Tao Wu}, {and} \bibinfo{person}{Xing Xie}.} \bibinfo{year}{2024}\natexlab{}.
\newblock \bibinfo{booktitle}{\emph{Recommender Systems: Frontiers and Practices}}.
\newblock \bibinfo{publisher}{Springer Nature}.
\newblock


\bibitem[Lichtenberg et~al\mbox{.}(2024)]%
        {lichtenberg2024large}
\bibfield{author}{\bibinfo{person}{Jan~Malte Lichtenberg}, \bibinfo{person}{Alexander Buchholz}, {and} \bibinfo{person}{Pola Schw{\"o}bel}.} \bibinfo{year}{2024}\natexlab{}.
\newblock \showarticletitle{Large language models as recommender systems: A study of popularity bias}.
\newblock \bibinfo{journal}{\emph{arXiv preprint arXiv:2406.01285}} (\bibinfo{year}{2024}).
\newblock


\bibitem[Liu et~al\mbox{.}(2023a)]%
        {liu2023recommendation}
\bibfield{author}{\bibinfo{person}{Jiahao Liu}, \bibinfo{person}{Dongsheng Li}, \bibinfo{person}{Hansu Gu}, \bibinfo{person}{Tun Lu}, \bibinfo{person}{Jiongran Wu}, \bibinfo{person}{Peng Zhang}, \bibinfo{person}{Li Shang}, {and} \bibinfo{person}{Ning Gu}.} \bibinfo{year}{2023}\natexlab{a}.
\newblock \showarticletitle{Recommendation unlearning via matrix correction}.
\newblock \bibinfo{journal}{\emph{arXiv preprint arXiv:2307.15960}} (\bibinfo{year}{2023}).
\newblock


\bibitem[Liu et~al\mbox{.}(2022)]%
        {liu2022parameter}
\bibfield{author}{\bibinfo{person}{Jiahao Liu}, \bibinfo{person}{Dongsheng Li}, \bibinfo{person}{Hansu Gu}, \bibinfo{person}{Tun Lu}, \bibinfo{person}{Peng Zhang}, {and} \bibinfo{person}{Ning Gu}.} \bibinfo{year}{2022}\natexlab{}.
\newblock \showarticletitle{Parameter-free dynamic graph embedding for link prediction}.
\newblock \bibinfo{journal}{\emph{Advances in Neural Information Processing Systems}}  \bibinfo{volume}{35} (\bibinfo{year}{2022}), \bibinfo{pages}{27623--27635}.
\newblock


\bibitem[Liu et~al\mbox{.}(2023b)]%
        {liu2023personalized}
\bibfield{author}{\bibinfo{person}{Jiahao Liu}, \bibinfo{person}{Dongsheng Li}, \bibinfo{person}{Hansu Gu}, \bibinfo{person}{Tun Lu}, \bibinfo{person}{Peng Zhang}, \bibinfo{person}{Li Shang}, {and} \bibinfo{person}{Ning Gu}.} \bibinfo{year}{2023}\natexlab{b}.
\newblock \showarticletitle{Personalized graph signal processing for collaborative filtering}. In \bibinfo{booktitle}{\emph{Proceedings of the ACM Web Conference 2023}}. \bibinfo{pages}{1264--1272}.
\newblock


\bibitem[Liu et~al\mbox{.}(2023c)]%
        {liu2023triple}
\bibfield{author}{\bibinfo{person}{Jiahao Liu}, \bibinfo{person}{Dongsheng Li}, \bibinfo{person}{Hansu Gu}, \bibinfo{person}{Tun Lu}, \bibinfo{person}{Peng Zhang}, \bibinfo{person}{Li Shang}, {and} \bibinfo{person}{Ning Gu}.} \bibinfo{year}{2023}\natexlab{c}.
\newblock \showarticletitle{Triple structural information modelling for accurate, explainable and interactive recommendation}. In \bibinfo{booktitle}{\emph{Proceedings of the 46th International ACM SIGIR Conference on Research and Development in Information Retrieval}}. \bibinfo{pages}{1086--1095}.
\newblock


\bibitem[Liu et~al\mbox{.}(2024)]%
        {liu2024filtering}
\bibfield{author}{\bibinfo{person}{Jiahao Liu}, \bibinfo{person}{Yiyang Shao}, \bibinfo{person}{Peng Zhang}, \bibinfo{person}{Dongsheng Li}, \bibinfo{person}{Hansu Gu}, \bibinfo{person}{Chao Chen}, \bibinfo{person}{Longzhi Du}, \bibinfo{person}{Tun Lu}, {and} \bibinfo{person}{Ning Gu}.} \bibinfo{year}{2024}\natexlab{}.
\newblock \showarticletitle{Filtering Discomforting Recommendations with Large Language Models}.
\newblock \bibinfo{journal}{\emph{arXiv preprint arXiv:2410.05411}} (\bibinfo{year}{2024}).
\newblock


\bibitem[Liu et~al\mbox{.}(2023d)]%
        {liu2023autoseqrec}
\bibfield{author}{\bibinfo{person}{Sijia Liu}, \bibinfo{person}{Jiahao Liu}, \bibinfo{person}{Hansu Gu}, \bibinfo{person}{Dongsheng Li}, \bibinfo{person}{Tun Lu}, \bibinfo{person}{Peng Zhang}, {and} \bibinfo{person}{Ning Gu}.} \bibinfo{year}{2023}\natexlab{d}.
\newblock \showarticletitle{Autoseqrec: Autoencoder for efficient sequential recommendation}. In \bibinfo{booktitle}{\emph{Proceedings of the 32nd ACM International Conference on Information and Knowledge Management}}. \bibinfo{pages}{1493--1502}.
\newblock


\bibitem[Madaan et~al\mbox{.}(2024)]%
        {madaan2024self}
\bibfield{author}{\bibinfo{person}{Aman Madaan}, \bibinfo{person}{Niket Tandon}, \bibinfo{person}{Prakhar Gupta}, \bibinfo{person}{Skyler Hallinan}, \bibinfo{person}{Luyu Gao}, \bibinfo{person}{Sarah Wiegreffe}, \bibinfo{person}{Uri Alon}, \bibinfo{person}{Nouha Dziri}, \bibinfo{person}{Shrimai Prabhumoye}, \bibinfo{person}{Yiming Yang}, {et~al\mbox{.}}} \bibinfo{year}{2024}\natexlab{}.
\newblock \showarticletitle{Self-refine: Iterative refinement with self-feedback}.
\newblock \bibinfo{journal}{\emph{Advances in Neural Information Processing Systems}}  \bibinfo{volume}{36} (\bibinfo{year}{2024}).
\newblock


\bibitem[Ortega et~al\mbox{.}(2024)]%
        {ortega2024evaluating}
\bibfield{author}{\bibinfo{person}{Gustavo~Mendon{\c{c}}a Ortega}, \bibinfo{person}{Rodrigo Ferrari~de Souza}, {and} \bibinfo{person}{Marcelo~Garcia Manzato}.} \bibinfo{year}{2024}\natexlab{}.
\newblock \showarticletitle{Evaluating zero-shot large language models recommenders on popularity bias and unfairness: a comparative approach to traditional algorithms}.
\newblock \bibinfo{journal}{\emph{Anais Estendidos}} (\bibinfo{year}{2024}).
\newblock


\bibitem[Pan et~al\mbox{.}(2023)]%
        {pan2023automatically}
\bibfield{author}{\bibinfo{person}{Liangming Pan}, \bibinfo{person}{Michael Saxon}, \bibinfo{person}{Wenda Xu}, \bibinfo{person}{Deepak Nathani}, \bibinfo{person}{Xinyi Wang}, {and} \bibinfo{person}{William~Yang Wang}.} \bibinfo{year}{2023}\natexlab{}.
\newblock \showarticletitle{Automatically correcting large language models: Surveying the landscape of diverse self-correction strategies}.
\newblock \bibinfo{journal}{\emph{arXiv preprint arXiv:2308.03188}} (\bibinfo{year}{2023}).
\newblock


\bibitem[Park et~al\mbox{.}(2023)]%
        {park2023generative}
\bibfield{author}{\bibinfo{person}{Joon~Sung Park}, \bibinfo{person}{Joseph O'Brien}, \bibinfo{person}{Carrie~Jun Cai}, \bibinfo{person}{Meredith~Ringel Morris}, \bibinfo{person}{Percy Liang}, {and} \bibinfo{person}{Michael~S Bernstein}.} \bibinfo{year}{2023}\natexlab{}.
\newblock \showarticletitle{Generative agents: Interactive simulacra of human behavior}. In \bibinfo{booktitle}{\emph{Proceedings of the 36th annual acm symposium on user interface software and technology}}. \bibinfo{pages}{1--22}.
\newblock


\bibitem[Petruzzelli et~al\mbox{.}(2024)]%
        {petruzzelli2024instructing}
\bibfield{author}{\bibinfo{person}{Alessandro Petruzzelli}, \bibinfo{person}{Cataldo Musto}, \bibinfo{person}{Lucrezia Laraspata}, \bibinfo{person}{Ivan Rinaldi}, \bibinfo{person}{Marco de Gemmis}, \bibinfo{person}{Pasquale Lops}, {and} \bibinfo{person}{Giovanni Semeraro}.} \bibinfo{year}{2024}\natexlab{}.
\newblock \showarticletitle{Instructing and prompting large language models for explainable cross-domain recommendations}. In \bibinfo{booktitle}{\emph{Proceedings of the 18th ACM Conference on Recommender Systems}}. \bibinfo{pages}{298--308}.
\newblock


\bibitem[Rendle et~al\mbox{.}(2012)]%
        {rendle2012bpr}
\bibfield{author}{\bibinfo{person}{Steffen Rendle}, \bibinfo{person}{Christoph Freudenthaler}, \bibinfo{person}{Zeno Gantner}, {and} \bibinfo{person}{Lars Schmidt-Thieme}.} \bibinfo{year}{2012}\natexlab{}.
\newblock \showarticletitle{BPR: Bayesian personalized ranking from implicit feedback}.
\newblock \bibinfo{journal}{\emph{arXiv preprint arXiv:1205.2618}} (\bibinfo{year}{2012}).
\newblock


\bibitem[Ricci et~al\mbox{.}(2010)]%
        {ricci2010introduction}
\bibfield{author}{\bibinfo{person}{Francesco Ricci}, \bibinfo{person}{Lior Rokach}, {and} \bibinfo{person}{Bracha Shapira}.} \bibinfo{year}{2010}\natexlab{}.
\newblock \showarticletitle{Introduction to recommender systems handbook}.
\newblock In \bibinfo{booktitle}{\emph{Recommender systems handbook}}. \bibinfo{publisher}{Springer}, \bibinfo{pages}{1--35}.
\newblock


\bibitem[Shani and Gunawardana(2011)]%
        {shani2011evaluating}
\bibfield{author}{\bibinfo{person}{Guy Shani} {and} \bibinfo{person}{Asela Gunawardana}.} \bibinfo{year}{2011}\natexlab{}.
\newblock \showarticletitle{Evaluating recommendation systems}.
\newblock \bibinfo{journal}{\emph{Recommender systems handbook}} (\bibinfo{year}{2011}), \bibinfo{pages}{257--297}.
\newblock


\bibitem[Shen et~al\mbox{.}(2024)]%
        {shen2024exploring}
\bibfield{author}{\bibinfo{person}{Tingjia Shen}, \bibinfo{person}{Hao Wang}, \bibinfo{person}{Jiaqing Zhang}, \bibinfo{person}{Sirui Zhao}, \bibinfo{person}{Liangyue Li}, \bibinfo{person}{Zulong Chen}, \bibinfo{person}{Defu Lian}, {and} \bibinfo{person}{Enhong Chen}.} \bibinfo{year}{2024}\natexlab{}.
\newblock \showarticletitle{Exploring User Retrieval Integration towards Large Language Models for Cross-Domain Sequential Recommendation}.
\newblock \bibinfo{journal}{\emph{arXiv preprint arXiv:2406.03085}} (\bibinfo{year}{2024}).
\newblock


\bibitem[Shi et~al\mbox{.}(2024)]%
        {shi2024large}
\bibfield{author}{\bibinfo{person}{Wentao Shi}, \bibinfo{person}{Xiangnan He}, \bibinfo{person}{Yang Zhang}, \bibinfo{person}{Chongming Gao}, \bibinfo{person}{Xinyue Li}, \bibinfo{person}{Jizhi Zhang}, \bibinfo{person}{Qifan Wang}, {and} \bibinfo{person}{Fuli Feng}.} \bibinfo{year}{2024}\natexlab{}.
\newblock \showarticletitle{Large language models are learnable planners for long-term recommendation}. In \bibinfo{booktitle}{\emph{Proceedings of the 47th International ACM SIGIR Conference on Research and Development in Information Retrieval}}. \bibinfo{pages}{1893--1903}.
\newblock


\bibitem[Shinn et~al\mbox{.}(2024)]%
        {shinn2024reflexion}
\bibfield{author}{\bibinfo{person}{Noah Shinn}, \bibinfo{person}{Federico Cassano}, \bibinfo{person}{Ashwin Gopinath}, \bibinfo{person}{Karthik Narasimhan}, {and} \bibinfo{person}{Shunyu Yao}.} \bibinfo{year}{2024}\natexlab{}.
\newblock \showarticletitle{Reflexion: Language agents with verbal reinforcement learning}.
\newblock \bibinfo{journal}{\emph{Advances in Neural Information Processing Systems}}  \bibinfo{volume}{36} (\bibinfo{year}{2024}).
\newblock


\bibitem[Shu et~al\mbox{.}(2024)]%
        {shu2024rah}
\bibfield{author}{\bibinfo{person}{Yubo Shu}, \bibinfo{person}{Haonan Zhang}, \bibinfo{person}{Hansu Gu}, \bibinfo{person}{Peng Zhang}, \bibinfo{person}{Tun Lu}, \bibinfo{person}{Dongsheng Li}, {and} \bibinfo{person}{Ning Gu}.} \bibinfo{year}{2024}\natexlab{}.
\newblock \showarticletitle{RAH! RecSys--Assistant--Human: A Human-Centered Recommendation Framework With LLM Agents}.
\newblock \bibinfo{journal}{\emph{IEEE Transactions on Computational Social Systems}} (\bibinfo{year}{2024}).
\newblock


\bibitem[Tang et~al\mbox{.}(2023)]%
        {tang2023one}
\bibfield{author}{\bibinfo{person}{Zuoli Tang}, \bibinfo{person}{Zhaoxin Huan}, \bibinfo{person}{Zihao Li}, \bibinfo{person}{Xiaolu Zhang}, \bibinfo{person}{Jun Hu}, \bibinfo{person}{Chilin Fu}, \bibinfo{person}{Jun Zhou}, {and} \bibinfo{person}{Chenliang Li}.} \bibinfo{year}{2023}\natexlab{}.
\newblock \showarticletitle{One model for all: Large language models are domain-agnostic recommendation systems}.
\newblock \bibinfo{journal}{\emph{arXiv preprint arXiv:2310.14304}} (\bibinfo{year}{2023}).
\newblock


\bibitem[Vajjala et~al\mbox{.}(2024)]%
        {vajjala2024cross}
\bibfield{author}{\bibinfo{person}{Ajay~Krishna Vajjala}, \bibinfo{person}{Dipak Meher}, \bibinfo{person}{Ziwei Zhu}, {and} \bibinfo{person}{David~S Rosenblum}.} \bibinfo{year}{2024}\natexlab{}.
\newblock \showarticletitle{Cross-Domain Recommendation Meets Large Language Models}.
\newblock \bibinfo{journal}{\emph{arXiv preprint arXiv:2411.19862}} (\bibinfo{year}{2024}).
\newblock


\bibitem[Wang et~al\mbox{.}(2024a)]%
        {wang2024survey}
\bibfield{author}{\bibinfo{person}{Lei Wang}, \bibinfo{person}{Chen Ma}, \bibinfo{person}{Xueyang Feng}, \bibinfo{person}{Zeyu Zhang}, \bibinfo{person}{Hao Yang}, \bibinfo{person}{Jingsen Zhang}, \bibinfo{person}{Zhiyuan Chen}, \bibinfo{person}{Jiakai Tang}, \bibinfo{person}{Xu Chen}, \bibinfo{person}{Yankai Lin}, {et~al\mbox{.}}} \bibinfo{year}{2024}\natexlab{a}.
\newblock \showarticletitle{A survey on large language model based autonomous agents}.
\newblock \bibinfo{journal}{\emph{Frontiers of Computer Science}} \bibinfo{volume}{18}, \bibinfo{number}{6} (\bibinfo{year}{2024}), \bibinfo{pages}{186345}.
\newblock


\bibitem[Wang et~al\mbox{.}(2024c)]%
        {wang2024user}
\bibfield{author}{\bibinfo{person}{Lei Wang}, \bibinfo{person}{Jingsen Zhang}, \bibinfo{person}{Hao Yang}, \bibinfo{person}{Zhi-Yuan Chen}, \bibinfo{person}{Jiakai Tang}, \bibinfo{person}{Zeyu Zhang}, \bibinfo{person}{Xu Chen}, \bibinfo{person}{Yankai Lin}, \bibinfo{person}{Hao Sun}, \bibinfo{person}{Ruihua Song}, {et~al\mbox{.}}} \bibinfo{year}{2024}\natexlab{c}.
\newblock \showarticletitle{User Behavior Simulation with Large Language Model-based Agents for Recommender Systems}.
\newblock \bibinfo{journal}{\emph{ACM Transactions on Information Systems}} (\bibinfo{year}{2024}).
\newblock


\bibitem[Wang et~al\mbox{.}(2023b)]%
        {wang2023rethinking}
\bibfield{author}{\bibinfo{person}{Xiaolei Wang}, \bibinfo{person}{Xinyu Tang}, \bibinfo{person}{Wayne~Xin Zhao}, \bibinfo{person}{Jingyuan Wang}, {and} \bibinfo{person}{Ji-Rong Wen}.} \bibinfo{year}{2023}\natexlab{b}.
\newblock \showarticletitle{Rethinking the evaluation for conversational recommendation in the era of large language models}.
\newblock \bibinfo{journal}{\emph{arXiv preprint arXiv:2305.13112}} (\bibinfo{year}{2023}).
\newblock


\bibitem[Wang et~al\mbox{.}(2023a)]%
        {wang2023recmind}
\bibfield{author}{\bibinfo{person}{Yancheng Wang}, \bibinfo{person}{Ziyan Jiang}, \bibinfo{person}{Zheng Chen}, \bibinfo{person}{Fan Yang}, \bibinfo{person}{Yingxue Zhou}, \bibinfo{person}{Eunah Cho}, \bibinfo{person}{Xing Fan}, \bibinfo{person}{Xiaojiang Huang}, \bibinfo{person}{Yanbin Lu}, {and} \bibinfo{person}{Yingzhen Yang}.} \bibinfo{year}{2023}\natexlab{a}.
\newblock \showarticletitle{Recmind: Large language model powered agent for recommendation}.
\newblock \bibinfo{journal}{\emph{arXiv preprint arXiv:2308.14296}} (\bibinfo{year}{2023}).
\newblock


\bibitem[Wang et~al\mbox{.}(2024b)]%
        {wang2024multi}
\bibfield{author}{\bibinfo{person}{Zhefan Wang}, \bibinfo{person}{Yuanqing Yu}, \bibinfo{person}{Wendi Zheng}, \bibinfo{person}{Weizhi Ma}, {and} \bibinfo{person}{Min Zhang}.} \bibinfo{year}{2024}\natexlab{b}.
\newblock \showarticletitle{Multi-Agent Collaboration Framework for Recommender Systems}.
\newblock \bibinfo{journal}{\emph{arXiv preprint arXiv:2402.15235}} (\bibinfo{year}{2024}).
\newblock


\bibitem[Xia et~al\mbox{.}(2022)]%
        {xia2022fire}
\bibfield{author}{\bibinfo{person}{Jiafeng Xia}, \bibinfo{person}{Dongsheng Li}, \bibinfo{person}{Hansu Gu}, \bibinfo{person}{Jiahao Liu}, \bibinfo{person}{Tun Lu}, {and} \bibinfo{person}{Ning Gu}.} \bibinfo{year}{2022}\natexlab{}.
\newblock \showarticletitle{FIRE: Fast incremental recommendation with graph signal processing}. In \bibinfo{booktitle}{\emph{Proceedings of the ACM Web Conference 2022}}. \bibinfo{pages}{2360--2369}.
\newblock


\bibitem[Yuan et~al\mbox{.}(2024)]%
        {yuan2024easytool}
\bibfield{author}{\bibinfo{person}{Siyu Yuan}, \bibinfo{person}{Kaitao Song}, \bibinfo{person}{Jiangjie Chen}, \bibinfo{person}{Xu Tan}, \bibinfo{person}{Yongliang Shen}, \bibinfo{person}{Ren Kan}, \bibinfo{person}{Dongsheng Li}, {and} \bibinfo{person}{Deqing Yang}.} \bibinfo{year}{2024}\natexlab{}.
\newblock \showarticletitle{Easytool: Enhancing llm-based agents with concise tool instruction}.
\newblock \bibinfo{journal}{\emph{arXiv preprint arXiv:2401.06201}} (\bibinfo{year}{2024}).
\newblock


\bibitem[Zang et~al\mbox{.}(2022)]%
        {zang2022survey}
\bibfield{author}{\bibinfo{person}{Tianzi Zang}, \bibinfo{person}{Yanmin Zhu}, \bibinfo{person}{Haobing Liu}, \bibinfo{person}{Ruohan Zhang}, {and} \bibinfo{person}{Jiadi Yu}.} \bibinfo{year}{2022}\natexlab{}.
\newblock \showarticletitle{A survey on cross-domain recommendation: taxonomies, methods, and future directions}.
\newblock \bibinfo{journal}{\emph{ACM Transactions on Information Systems}} \bibinfo{volume}{41}, \bibinfo{number}{2} (\bibinfo{year}{2022}), \bibinfo{pages}{1--39}.
\newblock


\bibitem[Zhang et~al\mbox{.}(2024c)]%
        {zhang2024generative}
\bibfield{author}{\bibinfo{person}{An Zhang}, \bibinfo{person}{Yuxin Chen}, \bibinfo{person}{Leheng Sheng}, \bibinfo{person}{Xiang Wang}, {and} \bibinfo{person}{Tat-Seng Chua}.} \bibinfo{year}{2024}\natexlab{c}.
\newblock \showarticletitle{On generative agents in recommendation}. In \bibinfo{booktitle}{\emph{Proceedings of the 47th international ACM SIGIR conference on research and development in Information Retrieval}}. \bibinfo{pages}{1807--1817}.
\newblock


\bibitem[Zhang et~al\mbox{.}(2024e)]%
        {zhang2024simulating}
\bibfield{author}{\bibinfo{person}{Guangping Zhang}, \bibinfo{person}{Dongsheng Li}, \bibinfo{person}{Hansu Gu}, \bibinfo{person}{Tun Lu}, \bibinfo{person}{Li Shang}, {and} \bibinfo{person}{Ning Gu}.} \bibinfo{year}{2024}\natexlab{e}.
\newblock \showarticletitle{Simulating News Recommendation Ecosystems for Insights and Implications}.
\newblock \bibinfo{journal}{\emph{IEEE Transactions on Computational Social Systems}} (\bibinfo{year}{2024}).
\newblock


\bibitem[Zhang et~al\mbox{.}(2024a)]%
        {zhang2024prospect}
\bibfield{author}{\bibinfo{person}{Jizhi Zhang}, \bibinfo{person}{Keqin Bao}, \bibinfo{person}{Wenjie Wang}, \bibinfo{person}{Yang Zhang}, \bibinfo{person}{Wentao Shi}, \bibinfo{person}{Wanhong Xu}, \bibinfo{person}{Fuli Feng}, {and} \bibinfo{person}{Tat-Seng Chua}.} \bibinfo{year}{2024}\natexlab{a}.
\newblock \showarticletitle{Prospect Personalized Recommendation on Large Language Model-based Agent Platform}.
\newblock \bibinfo{journal}{\emph{arXiv preprint arXiv:2402.18240}} (\bibinfo{year}{2024}).
\newblock


\bibitem[Zhang et~al\mbox{.}(2024d)]%
        {zhang2024agentcf}
\bibfield{author}{\bibinfo{person}{Junjie Zhang}, \bibinfo{person}{Yupeng Hou}, \bibinfo{person}{Ruobing Xie}, \bibinfo{person}{Wenqi Sun}, \bibinfo{person}{Julian McAuley}, \bibinfo{person}{Wayne~Xin Zhao}, \bibinfo{person}{Leyu Lin}, {and} \bibinfo{person}{Ji-Rong Wen}.} \bibinfo{year}{2024}\natexlab{d}.
\newblock \showarticletitle{Agentcf: Collaborative learning with autonomous language agents for recommender systems}. In \bibinfo{booktitle}{\emph{Proceedings of the ACM on Web Conference 2024}}. \bibinfo{pages}{3679--3689}.
\newblock


\bibitem[Zhang et~al\mbox{.}(2023)]%
        {zhang2023recommendation}
\bibfield{author}{\bibinfo{person}{Junjie Zhang}, \bibinfo{person}{Ruobing Xie}, \bibinfo{person}{Yupeng Hou}, \bibinfo{person}{Xin Zhao}, \bibinfo{person}{Leyu Lin}, {and} \bibinfo{person}{Ji-Rong Wen}.} \bibinfo{year}{2023}\natexlab{}.
\newblock \showarticletitle{Recommendation as instruction following: A large language model empowered recommendation approach}.
\newblock \bibinfo{journal}{\emph{ACM Transactions on Information Systems}} (\bibinfo{year}{2023}).
\newblock


\bibitem[Zhang et~al\mbox{.}(2021)]%
        {zhang2021cope}
\bibfield{author}{\bibinfo{person}{Yao Zhang}, \bibinfo{person}{Yun Xiong}, \bibinfo{person}{Dongsheng Li}, \bibinfo{person}{Caihua Shan}, \bibinfo{person}{Kan Ren}, {and} \bibinfo{person}{Yangyong Zhu}.} \bibinfo{year}{2021}\natexlab{}.
\newblock \showarticletitle{Cope: Modeling continuous propagation and evolution on interaction graph}. In \bibinfo{booktitle}{\emph{Proceedings of the 30th ACM International Conference on Information \& Knowledge Management}}. \bibinfo{pages}{2627--2636}.
\newblock


\bibitem[Zhang et~al\mbox{.}(2024b)]%
        {zhang2024survey}
\bibfield{author}{\bibinfo{person}{Zeyu Zhang}, \bibinfo{person}{Xiaohe Bo}, \bibinfo{person}{Chen Ma}, \bibinfo{person}{Rui Li}, \bibinfo{person}{Xu Chen}, \bibinfo{person}{Quanyu Dai}, \bibinfo{person}{Jieming Zhu}, \bibinfo{person}{Zhenhua Dong}, {and} \bibinfo{person}{Ji-Rong Wen}.} \bibinfo{year}{2024}\natexlab{b}.
\newblock \showarticletitle{A survey on the memory mechanism of large language model based agents}.
\newblock \bibinfo{journal}{\emph{arXiv preprint arXiv:2404.13501}} (\bibinfo{year}{2024}).
\newblock


\bibitem[Zhang et~al\mbox{.}(2024f)]%
        {zhang2024llm}
\bibfield{author}{\bibinfo{person}{Zijian Zhang}, \bibinfo{person}{Shuchang Liu}, \bibinfo{person}{Ziru Liu}, \bibinfo{person}{Rui Zhong}, \bibinfo{person}{Qingpeng Cai}, \bibinfo{person}{Xiangyu Zhao}, \bibinfo{person}{Chunxu Zhang}, \bibinfo{person}{Qidong Liu}, {and} \bibinfo{person}{Peng Jiang}.} \bibinfo{year}{2024}\natexlab{f}.
\newblock \showarticletitle{LLM-Powered User Simulator for Recommender System}.
\newblock \bibinfo{journal}{\emph{arXiv preprint arXiv:2412.16984}} (\bibinfo{year}{2024}).
\newblock


\bibitem[Zhao et~al\mbox{.}(2023)]%
        {zhao2023survey}
\bibfield{author}{\bibinfo{person}{Wayne~Xin Zhao}, \bibinfo{person}{Kun Zhou}, \bibinfo{person}{Junyi Li}, \bibinfo{person}{Tianyi Tang}, \bibinfo{person}{Xiaolei Wang}, \bibinfo{person}{Yupeng Hou}, \bibinfo{person}{Yingqian Min}, \bibinfo{person}{Beichen Zhang}, \bibinfo{person}{Junjie Zhang}, \bibinfo{person}{Zican Dong}, {et~al\mbox{.}}} \bibinfo{year}{2023}\natexlab{}.
\newblock \showarticletitle{A survey of large language models}.
\newblock \bibinfo{journal}{\emph{arXiv preprint arXiv:2303.18223}} (\bibinfo{year}{2023}).
\newblock


\bibitem[Zhao et~al\mbox{.}(2024)]%
        {zhao2024let}
\bibfield{author}{\bibinfo{person}{Yuyue Zhao}, \bibinfo{person}{Jiancan Wu}, \bibinfo{person}{Xiang Wang}, \bibinfo{person}{Wei Tang}, \bibinfo{person}{Dingxian Wang}, {and} \bibinfo{person}{Maarten de Rijke}.} \bibinfo{year}{2024}\natexlab{}.
\newblock \showarticletitle{Let me do it for you: Towards llm empowered recommendation via tool learning}. In \bibinfo{booktitle}{\emph{Proceedings of the 47th International ACM SIGIR Conference on Research and Development in Information Retrieval}}. \bibinfo{pages}{1796--1806}.
\newblock


\bibitem[Zhu et~al\mbox{.}(2024a)]%
        {zhu2024reliable}
\bibfield{author}{\bibinfo{person}{Lixi Zhu}, \bibinfo{person}{Xiaowen Huang}, {and} \bibinfo{person}{Jitao Sang}.} \bibinfo{year}{2024}\natexlab{a}.
\newblock \showarticletitle{How Reliable is Your Simulator? Analysis on the Limitations of Current LLM-based User Simulators for Conversational Recommendation}. In \bibinfo{booktitle}{\emph{Companion Proceedings of the ACM on Web Conference 2024}}. \bibinfo{pages}{1726--1732}.
\newblock


\bibitem[Zhu et~al\mbox{.}(2024b)]%
        {zhu2024llm}
\bibfield{author}{\bibinfo{person}{Lixi Zhu}, \bibinfo{person}{Xiaowen Huang}, {and} \bibinfo{person}{Jitao Sang}.} \bibinfo{year}{2024}\natexlab{b}.
\newblock \showarticletitle{A LLM-based Controllable, Scalable, Human-Involved User Simulator Framework for Conversational Recommender Systems}.
\newblock \bibinfo{journal}{\emph{arXiv preprint arXiv:2405.08035}} (\bibinfo{year}{2024}).
\newblock


\end{thebibliography}

\appendix

\end{document}